\def\erg{{\rm\thinspace erg}}

\def\Msun{\hbox{$\rm\thinspace M_{\odot}$}}

\def\s{{\rm\thinspace s}}
\def\yr{{\rm\thinspace yr}}

\def\ergps{\hbox{$\erg\s^{-1}\,$}}

\documentclass{mn2e}

\usepackage{mathptmx}
\usepackage{graphicx}

\title[X-ray sources in the HVS of NGC\,1275]
{The Ultra Luminous X-ray sources in the High Velocity System of NGC\,1275}
\author[O. Gonz\'alez-Mart\'in, A.C. Fabian and J.S. Sanders]
{O. Gonz\'alez-Mart\'in$^{1,2}$\thanks{E-mail: omaira@iaa.es},
A.C. Fabian$^2$ and J.S. Sanders$^2$\\
\\ $^1$Instituto de Astrof\'isica de Andaluc\'ia (CSIC), Apdo 3004, 18080 Granada, Spain
\\ $^2$Institute of Astronomy, Madingley Road, Cambridge CB3 0HA
}

\begin{document}

\maketitle

\label{firstpage}

\begin{abstract} We report the results of a study of X-ray point
sources coincident with the High Velocity System (HVS) projected in
front of NGC\,1275. A very deep X-ray image of the core of the 
Perseus cluster made with the \emph{Chandra Observatory} has been
used. We find a population of Ultra-Luminous X-ray sources (ULX; 7
sources with L$_{\rm X}(0.5-7.0 \rm{\,keV})>7\times 10^{39}~ \rm{erg ~
s^{-1}}$).  As with the ULX populations in the Antennae and Cartwheel
galaxies, those in the HVS are associated with a region of very active
star formation. Several sources have possible optical counterparts
found on \emph{HST} images, although the X-ray brightest one does not.
Absorbed power-law models fit the X-ray spectra, with most having a
photon index between 2 and 3. 
\end{abstract} 

\begin{keywords}
galaxies: clusters: individual: Perseus - ULX - galaxies: individual:
NGC\,1275 
\end{keywords}

\section{Introduction}

\begin{figure*}
 \includegraphics[width=0.8\textwidth]{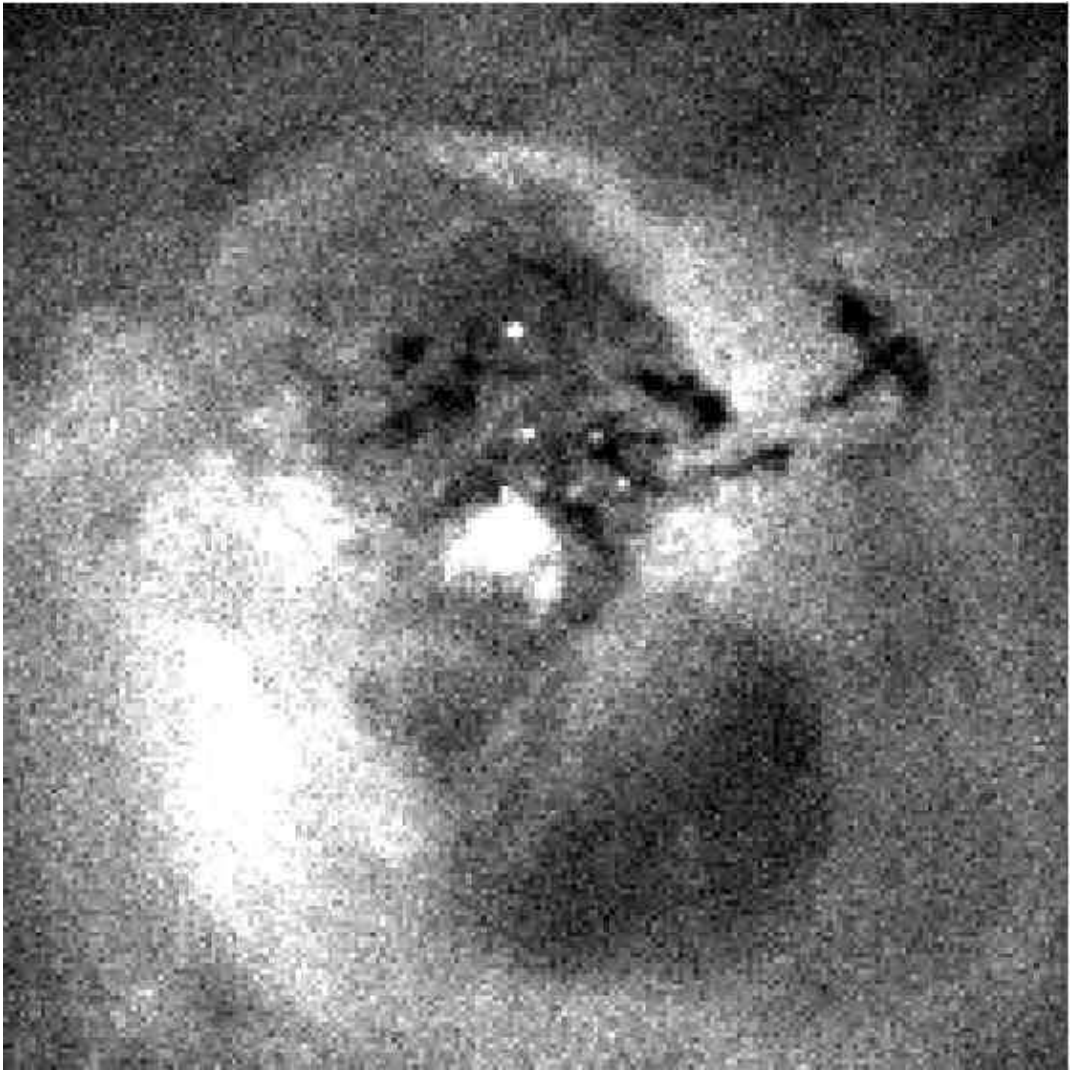}
 \caption{NGC\,1275 in the 0.3--7.0 keV band. Pixels are 0.49 arcsec
   in size and the N-S height of the image is 97~arcsec.  North is to
   the top and east is to the left in this image.  The point sources
   which we identify as ULX are located in the high-velocity system is
   seen in absorption to the north of the bright nucleus. None are
   seen in the south lobe}\label{fig:whole_picture}
\end{figure*}
\begin{figure*}
  \includegraphics[width=0.8\textwidth]{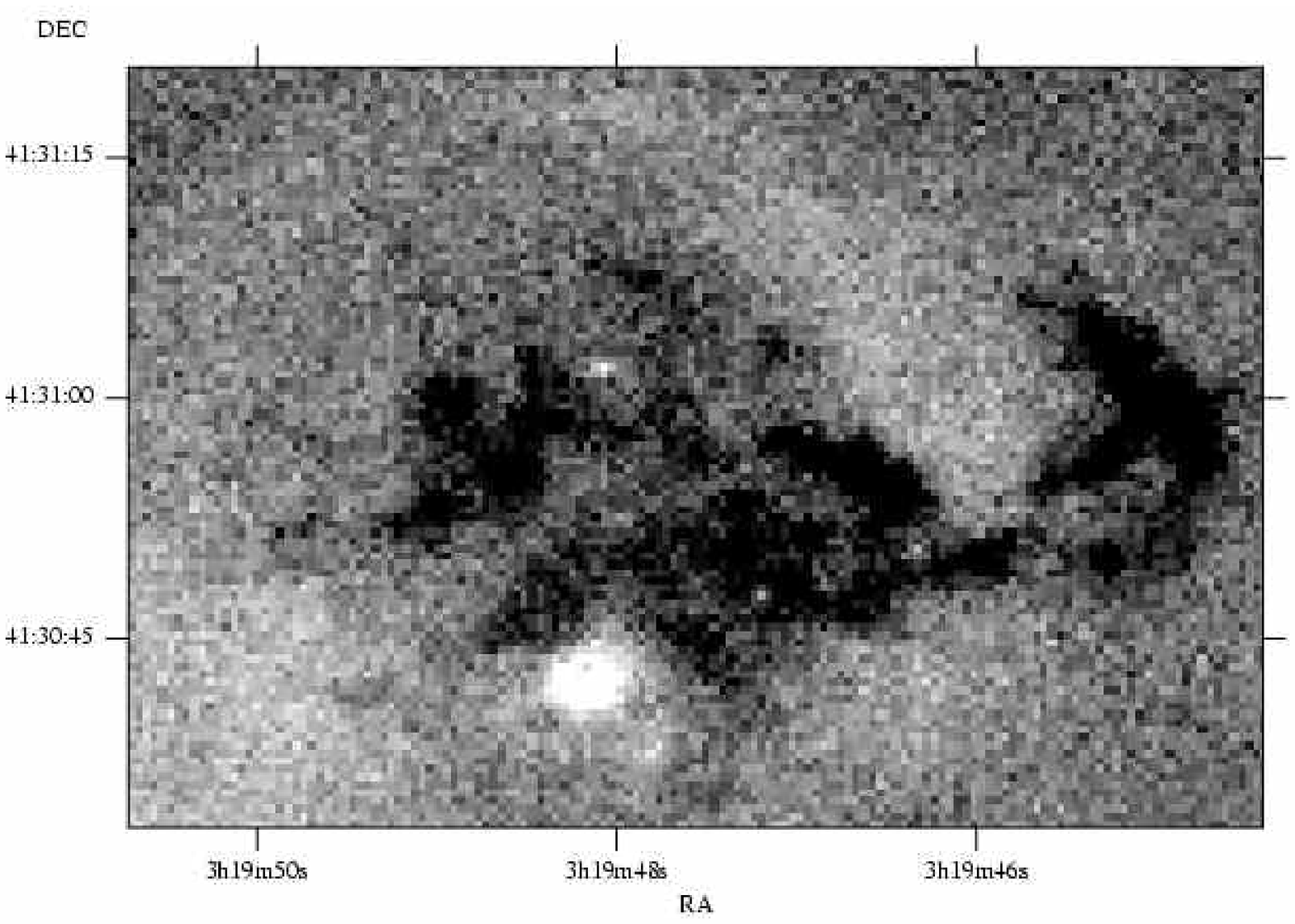}
  \caption{The central region of NGC\,1275 in the 0.3--0.8 keV band. Pixels are
    0.49 arcsec in dimension and the entire image is $1.77\times 0.89
    ~ \rm{arcmin^{2}}$.
    North is to the top and east is to the left in this image. The high-velocity
    system is seen in absorption to the north of the bright nucleus
    which is at RA $3^{\rm h}
    19^{\rm m} 48^{\rm s}$, Dec. +$41^o$30'42"}\label{fig:HVS}
\end{figure*}

The study of Ultra-Luminous X-ray sources (ULX) has been greatly
expanded by the high spatial resolution and spectral grasp of the
\emph{Chandra} and \emph{XMM-Newton} observatories, respectively. ULX
sources (Fabbiano \& White 2003; Miller \& Colbert 2004) have
2--10~keV X-ray luminosities exceeding $10^{39}\ergps$ and are found
some distance from the centres of galaxies; they are not active
galactic nuclei.  Their luminosity exceeds that for a $10\Msun$ black
hole accreting at the Eddington limit which radiates isotropically and
so have created much interest in the possibility that they contain
even higher mass black holes, such as InterMediate Black Holes (IMBH)
of $\sim10^3\Msun$ (Makishima et al 2000; Miller, Fabian \& Miller 2004).
Alternatively they may appear so luminous because of beaming (Reynolds
et al 1999; King et al 2001, Zezas \& Fabbiano 2002) or due to super
Eddington accretion (Begelman 2002).

ULX are most common in starburst galaxies and in very active
star-forming regions, such as in the Antennae and the Cartwheel
galaxy, where populations of tens of them are found (Zezas et al 2002;
Gao et al 2003; Wolter \& Trinchieri 2004). In some cases variability
rules out the possibility that they are just clusters of
lower-luminosity objects. The origin of IMBH is unclear. They may form
as a result of binary interactions in dense stellar environments
(Portegies Zwart \& McMillan 2002).  A comparison of IMBH ULX
candidates with a number of well known stellar-mass black holes
candidates (BHC; Miller et al 2004) demonstrates that the
ULX are more luminous but have cooler thermal disk components than
standard stellar-mass BHC.  Therefore, ULX in this sample are clearly
different from the sample of stellar-mass BHC and are consistent with
being IMBH.

Here we report on the discovery of a population of 8 point X-ray
sources to the N of the nucleus of NGC\,1275, which is the central
galaxy in the Perseus cluster. All exceed $10^{39}\ergps$ in X-ray
luminosity, and 7 are formally ULX, if they are at the distance of the
cluster. The spatial region where they lie coincides with the High
Velocity System of NGC\,1275. We assume that they are part of that
system. We see no other point sources (apart from the nucleus) 
over the body of NGC\,1275
(Fig.~1).

NGC\,1275 is embedded in a complex multiphase environment. Optical
imaging and spectroscopy first established the existence of two
distinct emission-line system toward NGC\,1275: a low-velocity
component associated with the galaxy itself at 5200 km $\rm{s^{-1}}$
and a high-velocity component at 8200 km $\rm{s^{-1}}$ projected
nearby on the sky (Minkowski 1955, 1957). This latter component is
associated with a small gas-rich galaxy falling into the cluster along
our line of sight (Haschick, Crane \& van der Hulst 1982). A merger
scenario has been proposed (Minkowski 1955, 1957). However,
interaction of the low and/or high-velocity system with a third
gas-rich galaxy or system of galaxies (Holtzman et al. 1992;
Conselice, Gallagher \& Wyse 2001), or influences from the surrounding
dense intracluster medium (ICM) (Sarazin 1988; Boroson 1990; Caulet et
al. 1992) have been discussed.

Deep \emph{Chandra} observations have clarified the position of the
High Velocity System (HVS).  The depth of the observed X-ray
absorption (e.g. Fig.~1) is nor infilled by emission from hot gas
projected along the line-of-sight so the HVS must lie well in front of
NGC\,1275. Gillmon, Sanders \& Fabian (2004) have estimated a lower
limit on the distance of the HVS from the nucleus of 57 kpc.  The low-
and high-velocity system are therefore not yet directly interacting.
The HVS $\emph{is}$ however strongly interacting with the ICM of the
Perseus cluster, which has triggered strong star formation.  In this
paper we describe the detailed analysis of the X-ray spatial and
spectral properties of the discrete sources in the high velocity
system.

The paper is organized as follows: in Sect. 2 and 3 we present
reduction and results from the imaging analysis and spectral analysis,
respectively; in Sect. 4 we discuss the results; and Sect. 5
summarizes our findings.

Throughout this paper we use a redshift of 0.018 and
$H_0=70~\rm{km~s^{-1}~Mpc^{-1}}$. This gives a luminosity distance to
the cluster of 80 Mpc; 1 arcsec corresponds to a physical distance of
370~pc.

\section{Imaging analysis}
The \emph{Chandra} datasets included in this analysis are listed in
Table~\ref{tab:obs}. The total exposure time, after removing periods
containing flares, is 890~ks. To prepare the data for analysis, all of
the datasets were reprocessed to use the latest appropriate gain file
(acisD2000-01-29gain\_ctiN0003). The datasets analysed each used an
aimpoint on the ACIS-S3 CCD.  The datasets were filtered using the
lightcurve in the 2.5 to 7~keV band on ACIS-S1 CCD, which is a
back-illuminated CCD like the ACIS-S3. The CIAO \textsc{lc\_clean}
tool was used to remove periods 20 per~cent away from the median count
rate for all the lightcurves. This procedure was not used for datasets
03209 and 04289 which did not include the S1 CCD, however no flares
were seen in these observations on the S3 CCD. Each of the
observations was reprojected to match the coordinate system of the
04952 observation.

\begin{table*}
  \begin{tabular}{lllllll}
    Obs. ID & Sequence & Observation date & Exposure (ks) & Nominal roll
    (deg) & Pointing RA & Pointing Dec \\ \hline
    3209 & 800209 & 2002-08-08 &  95.8 & 101.2 & 3:19:46.86 & +41:31:51.3 \\
    4289 & 800209 & 2002-08-10 &  95.4 & 101.2 & 3:19:46.86 & +41:31:51.3 \\
    6139 & 800397 & 2004-10-04 &  51.6 & 125.9 & 3:19:45.54 & +41:31:33.9 \\
    4946 & 800397 & 2004-10-06 &  22.7 & 127.2 & 3:19:45.44 & +41:31:33.2 \\
    4948 & 800398 & 2004-10-09 & 107.5 & 128.9 & 3:19:44.75 & +41:31:40.1 \\
    4947 & 800397 & 2004-10-11 &  28.7 & 130.6 & 3:19:45.17 & +41:31:31.3 \\
    4949 & 800398 & 2004-10-12 &  28.8 & 130.9 & 3:19:44.57 & +41:31:38.7 \\
    4950 & 800399 & 2004-10-12 &  73.4 & 131.1 & 3:19:43.97 & +41:31:46.1 \\
    4952 & 800400 & 2004-10-14 & 143.2 & 132.6 & 3:19:43.22 & +41:31:52.2 \\
    4951 & 800399 & 2004-10-17 &  91.4 & 135.2 & 3:19:43.57 & +41:31:42.6 \\
    4953 & 800400 & 2004-10-18 &  29.3 & 136.2 & 3:19:42.83 & +41:31:48.5 \\
    6145 & 800397 & 2004-10-19 &  83.1 & 137.7 & 3:19:44.66 & +41:31:26.7 \\
    6146 & 800398 & 2004-10-20 &  39.2 & 138.7 & 3:19:43.92 & +41:31:32.7 \\
  \end{tabular}
  \caption{\emph{Chandra} observations included in this analysis. The
    exposure given is the time remaining after filtering the
    lightcurve for flares. All observations were taken with the
    aimpoint on the ACIS-S3 CCD. All positions are in J2000 coordinates.}
  \label{tab:obs}
\end{table*}

The 900~ks X-ray image covering the energy range $\rm{0.3-0.8~keV}$ is
shown in Fig \ref{fig:HVS}. The bright NGC\,1275 nucleus is clearly
seen at RA $3^h 19^m 48^s$ and Dec. +$41^o$30'42" (J2000) and the
high-velocity system is seen in absorption to the north of the
nucleus.

The CIAO \textsc{celldetect} source detection routine was then used on
the reprocessed level 2 event data to produce a preliminary list of
point sources. The cell size ranges between 4 pixels to 8 pixels. This
algorithm strongly depends on the local background and the detection
cell in not adjustable to the size of the source. As the X-ray diffuse
emission of the NGC 1275 is very strong, the source list may well
include false detections in high background level regions. Therefore
problematic sources embedded in such regions have been excluded in our
analysis.  Moreover, as mentioned above, we only included sources
associated with the HVS.

We have detected 8 bright sources close to the nucleus of NGC\,1275,
located in the northern inner radio lobe of 3C 84. All of these source
are embedded in the same region as the HVS (see Fig.
\ref{fig:whole_picture}).  There are no sources associated with the
southern lobe (Fig. \ref{fig:whole_picture}), thus we assume these
sources are associated with the HVS.

Fig. \ref{fig:figure} (\emph{left}) shows the smoothed ACIS-S3 image in the
0.3--7.0 keV band, including numbered labels of all the detected sources
(\emph{top}), centred on source labelled N3 (\emph{centre}) and centred 
on source N5 (\emph{bottom}).

All the point-like sources are listed in Table~\ref{tab:positions},
showing their positions and count rates.

 \begin{table}
\begin{center}
  \begin{tabular}{lcc} \hline \hline 
N & Position(J2000)& 	Count Rate	  \\
   &                &    (counts $\rm{ ks^{-1}}$)           \\ \hline  
1.... &  03:19:48.736    +41:30:47.25 &	0.34$\pm$0.05   \\   
2.... &  03:19:48.166    +41:30:46.64 & 1.69$\pm$0.06 \\  
3.... &  03:19:48.090    +41:31:01.88 & 2.60$\pm$0.08  \\  
4.... &  03:19:47.994    +41:30:52.30 & 1.42$\pm$0.09\\  
5.... &  03:19:47.925    +41:30:47.50 & 1.19$\pm$0.09 \\  
6.... &  03:19:47.602    +41:30:47.01 & 0.74$\pm$0.06 \\  
7.... &  03:19:47.422    +41:30:51.93 & 0.95$\pm$0.08\\  
8.... &  03:19:47.214    +41:30:47.62 & 1.28$\pm$0.08\\  \hline  
\end{tabular} 
  \caption{Positions of sources detected near the NGC\,1275 centre and
  displayed in Fig. \ref{fig:figure} (column 2) and count rate in the 
  energy range between 0.5--7.0 keV (column 3).} 
  \label{tab:positions}
  \end{center}
\end{table}

We have used archival \emph{HST} observations of NGC\,1275 in order to
search for optical counterparts. The galaxy was imaged with the WFPC2
camera on \emph{HST} using the F814W ($\sim$ I, on 2001 November 6
with an exposure time of 1200~s) and F702W ($\sim$ R, on 1994 March 31
with an exposure time of 140~s) broad-band filters. Several
coincidences between X-ray sources and optical knots of emission
(F814W) can be seen in Fig. \ref{fig:figure} ({right}), showing the
same regions as Fig. \ref{fig:figure} ({left}).

The \emph{HST} image shows many highly absorbed features. When we
compare in detail, sources N7 and N8 are located in star forming
regions, while N2 and N6 have a point-like counterpart. Sources N1,
N3, N4 and N5 have no optical identification. Therefore, we have found
a possible correlation between compact X-ray sources and regions of
vigorous star formation. The implications are discussed later.

 \begin{table}
\begin{center}
  \begin{tabular}{lcccc} \hline \hline 
N 	& F${\rm _X}$/F$_{{\rm F814W}}$& F$_{\rm X}$/F$_{{\rm
F702W}}$& ${\rm M_{F814W}}$ & ${\rm M_{F702W}}$	 \\
   	   &                &    		&               &\\ \hline  
       1....   &   $>26.5$		&    $>16.2$	    &	 $>$22.6    &	$>$22.1      \\
       2....   &   25.4 	&    23.6	    &	 20.4	    &	20.3	     \\
       3....   &   $>$18800	&    ...	    &	 $>$26.8    &	...	     \\
       4....   &   $>$1081	&    $>$800	    &	 $>$24.5    &	$>$24.2      \\
       5....   &   $>$123	&    $>$60.6	    &	 $>$22.6    &	$>$21.9      \\
       6....   &   26.2 	&    28.4	    &	 21.7	    &	21.7	     \\
       7....   &   76.7 	&    51.4	    &	 22.3	    &	21.9	     \\
       8....   &   134  	&    90.1	    &	 22.2	    &	21.7	     \\  \hline 
   \end{tabular}
  \caption{Optical analysis. X-ray to optical ratios (columns 2 and 3) and
  magnitude determinations (columns 4 and 5) for the filters F814W  and F702W,
  respectively, with the X-ray flux between 1.0--7.0~keV.} 
  \label{tab:optical}
  \end{center}
\end{table}

In order to investigate the emission mechanism of these ULX, the X-ray
to optical flux ratios have been computed between the F702W and F814W
\emph{HST} broad-bands and 1.0--7.0~keV X-ray band.  Preliminary
processing of the raw images including corrections for the flat
fielding was done remotely at the \emph{Space Telescope Science
  Institute} through the standard pipeline. For each frame, cosmic
rays were removed by image combination, using the {\sc imcombine}
routine in IRAF. After cosmic ray removal, the frames were added using
task {\sc wmosaic} in STSDAS package. Photometric measurements were
made with {\sc phot} task, within the NOAO package. Finally the fluxes
and magnitudes have been determined using the photometric zero-point
information in the header of the calibrated image files.

These results are shown in Table \ref{tab:optical}, including the
X-ray to optical flux ratios from the F814W and F702W broad-band
filters, and the magnitude determinations from the same filters.  In
the cases where an optical counterpart has not been found (N1, N3, N4
and N5), the magnitudes and fluxes are just a lower limit.

\section{Spectral analysis}
We extracted spectra for all the detected sources close to the HVS,
using extraction regions defined to include as many of the
source photons as possible, but at the same time minimizing
contamination from nearby sources and background. The background
region was either a source-free circular annulus or several circles
surrounding each source, in order to take into account the spatial
variations of the diffuse emission and to minimize effects related to
the spatial variation of the CCD response.

For each source, we extracted spectra from each of the datasets. These
spectra were summed to form a total spectrum for each source. Response
and ancillary response files were created for each source in each of
the observations using the CIAO \textsc{mkacisrmf} and \textsc{mkwarf}
tools. The responses for a particular source were summed together,
weighting according to the number of counts in each observation.

The spectra were fitted using XSPEC v.11.3.2. In order to use the
${\rm \chi^{2}}$ statistic, we grouped the data to include at least 20
counts per spectral bin, before background subtraction. In spectral
fitting we excluded any events with energies above 7.0 keV or below
0.5 keV.

 \begin{table}
 \begin{center}
  \begin{tabular}{lccr} \hline \hline 
N &  N${_{\rm H}}$		  &	${\rm \Gamma}$	   &	${\rm \chi^{2}}$/d.o.f.	       \\ 
  & (${\rm 10^{21}cm^{-2}}$)    & 	   &   \\ \hline  
1.... &   2.5$^{(a)}$ 	             & 3.20$^{+0.23}_{-0.37}$   & 112.90/101  \\  
2.... &   2.72$^{+1.43}_{-0.87}$     & 1.78$^{+0.30}_{-0.24}$		& 101.50/109	    \\  
3.... &   2.49$^{+0.40}_{-0.40}$     & 2.08$^{+0.09}_{-0.09}$ &  153.86/142 \\  
4.... &   2.05$^{+0.91}_{-0.96}$     & 2.29$^{+0.44}_{-0.28}$	&  156.24/152	    \\  
5.... &   2.64$^{+1.23}_{-0.93}$     & 2.92$^{+1.44}_{-0.36}$	&  124.09/139	    \\  
6.... &   3.74$^{+1.57}_{-1.39}$     & 3.51$^{+0.48}_{-0.66}$	&  102.58/92   \\  
7.... &   4.03$^{+1.78}_{-1.45}$     & 3.20$^{+1.39}_{-0.48}$	&  133.69/135	    \\  
8.... &   2.66$^{+1.00}_{-0.91}$     & 2.13$^{+0.52}_{-0.25}$	&  150.81/138	    \\  \hline   
\end{tabular}
  \caption{Spectral fits. (a) The column density of source N1 has been fixed due to the low 
  count rate.} 
  \label{tab:fittings}
  \end{center}
\end{table}

Table \ref{tab:fittings} summarizes our spectral results in terms of
the absorbing column density and photon index.

The sources have been modelled with an absorbed power law slope with
photon index between ${\rm \Gamma=}$[1.78-5.56] and an equivalent
column density of $\rm{N_H=[2.05-4.03]\times 10^{21} cm^{-2}}$. In all
the cases the single component power law give satisfactory fits. The
column density of source N1 has been fixed due to the low count rate.
The fitted $\rm{N_H}$ values are consistent with the intrinsic
absorption measured e.g. in the optical band; the value of
A$\rm{_V}$=0.54 corresponds to $\rm{N_H\sim 1.1\times
10^{21}cm^{-2}}$, assuming $\rm{A_V=N_H \times 5.3 \times 10^{-22}}$
for $\rm{ R_V=3.1}$ (Bohlin et al. 1978). This value should be a lower
limit to the fitted $\rm{N_H}$ value to be consistent, as is seen in
Table \ref{tab:fittings}.

\begin{figure}
  \includegraphics[width=0.65\columnwidth,angle=-90]{source1_spec.eps}
  \caption{ACIS-S spectrum  of source N3. The solid line corresponds with a
    power law model with a spectral index of $\rm{\Gamma=}$2.08 and absorption of 
    $\rm{N_H=2.5 \times 10^{21}~cm^{-2}}$. The fit residuals are
    presented in the lower panel.}
  \label{fig:source1_spec}
\end{figure}

\begin{figure*}
 \includegraphics[width=\columnwidth]{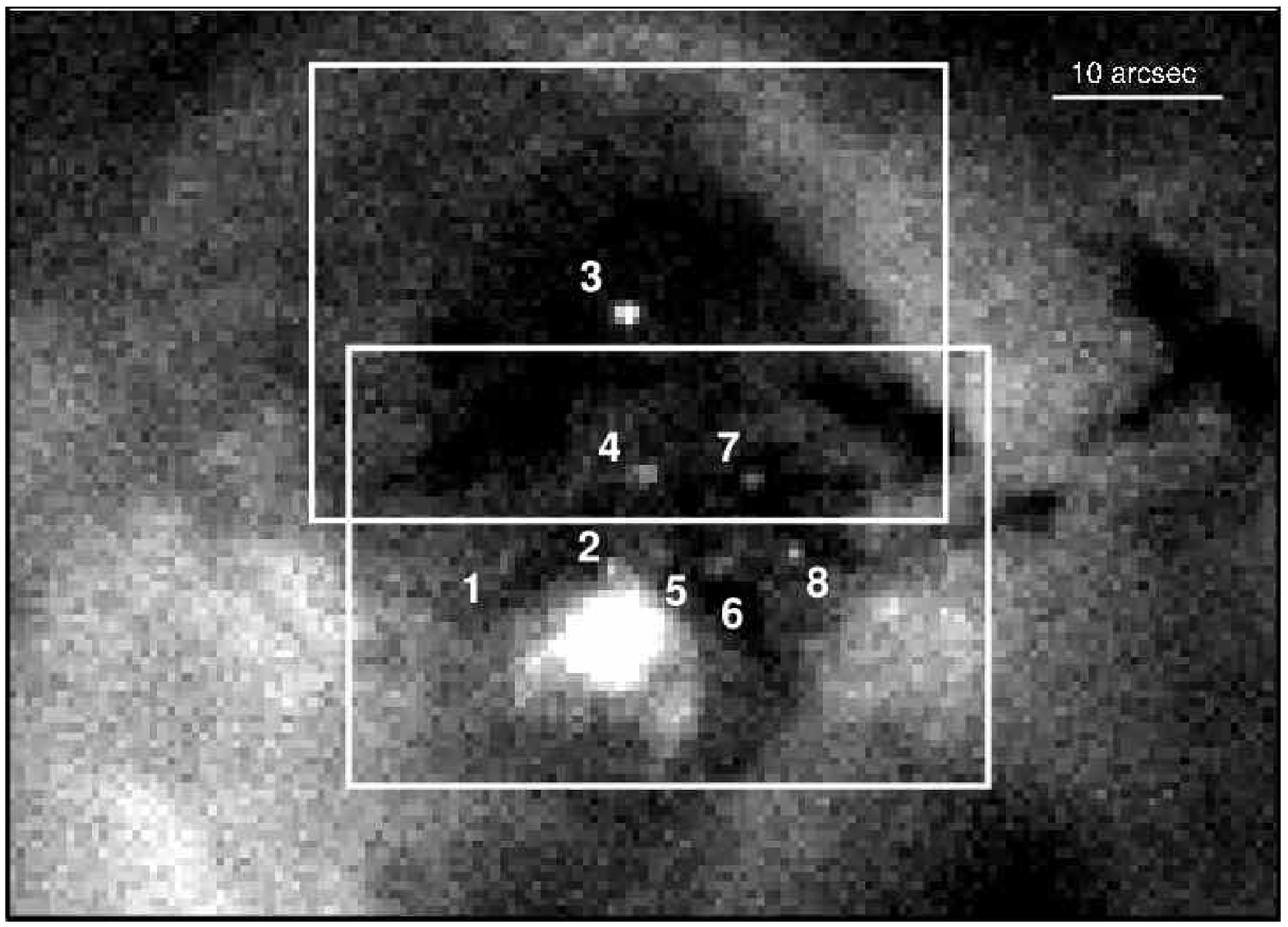}
 \includegraphics[width=\columnwidth]{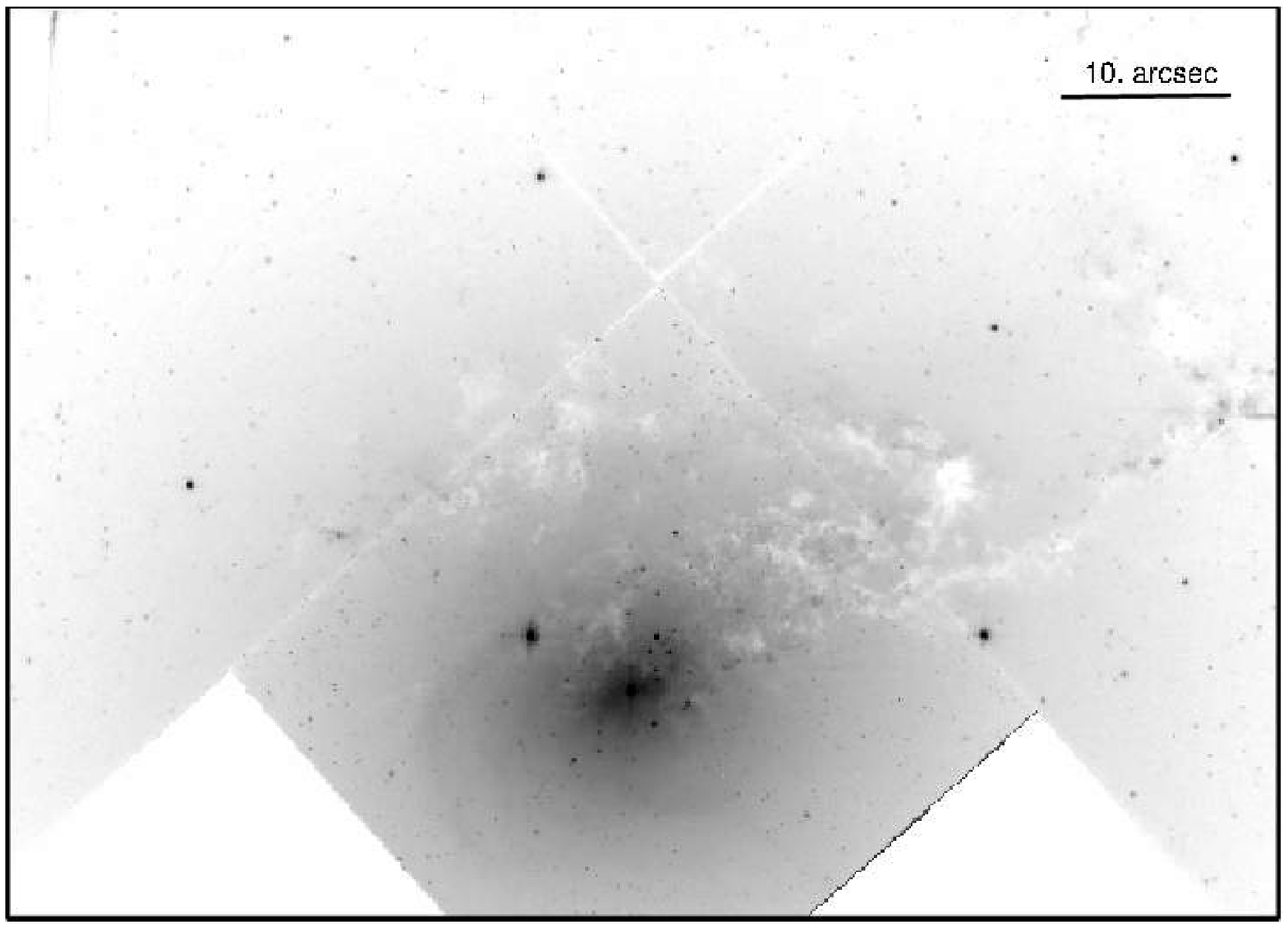}
 \includegraphics[width=\columnwidth]{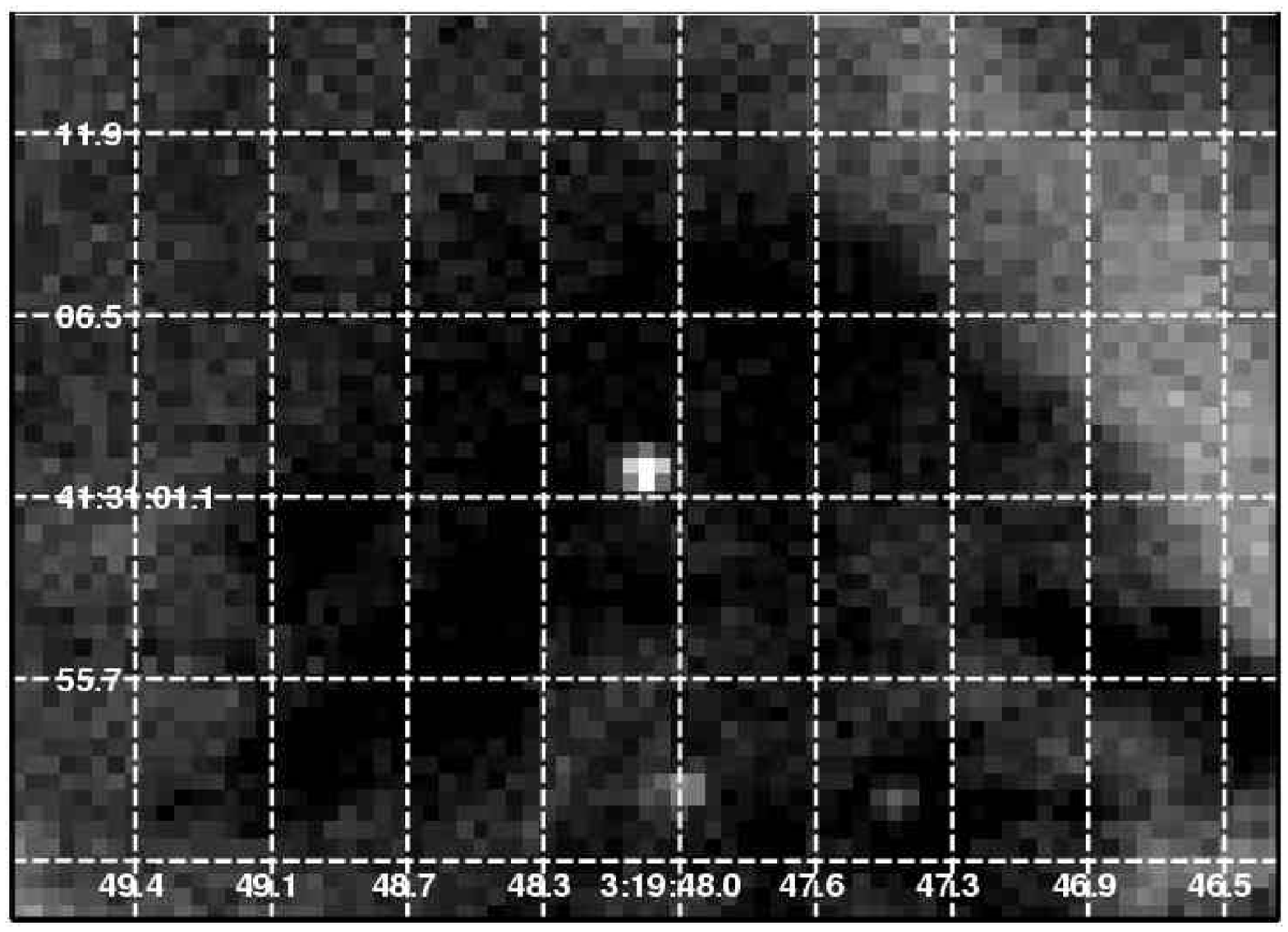}
 \includegraphics[width=\columnwidth]{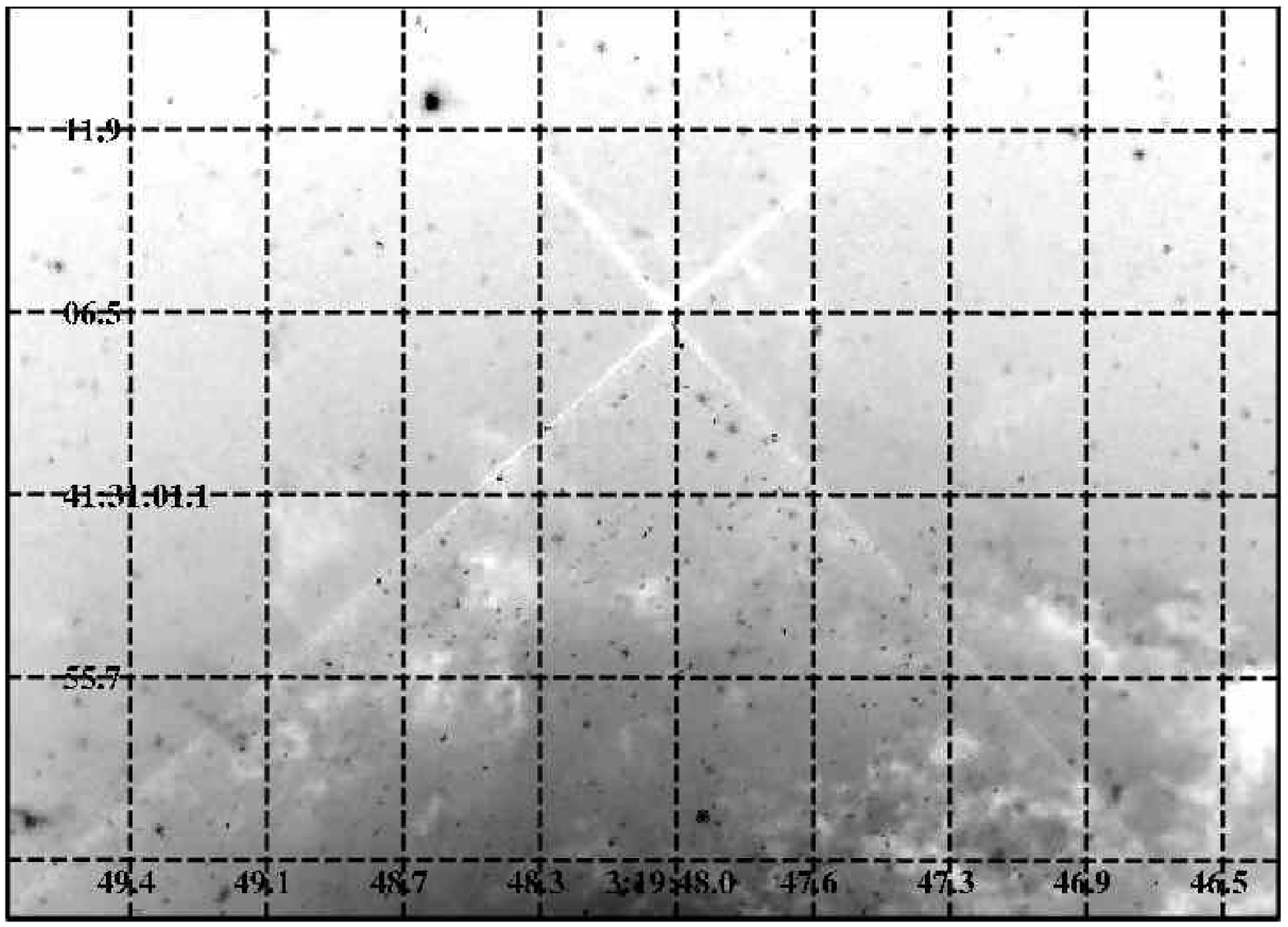}
 \includegraphics[width=\columnwidth]{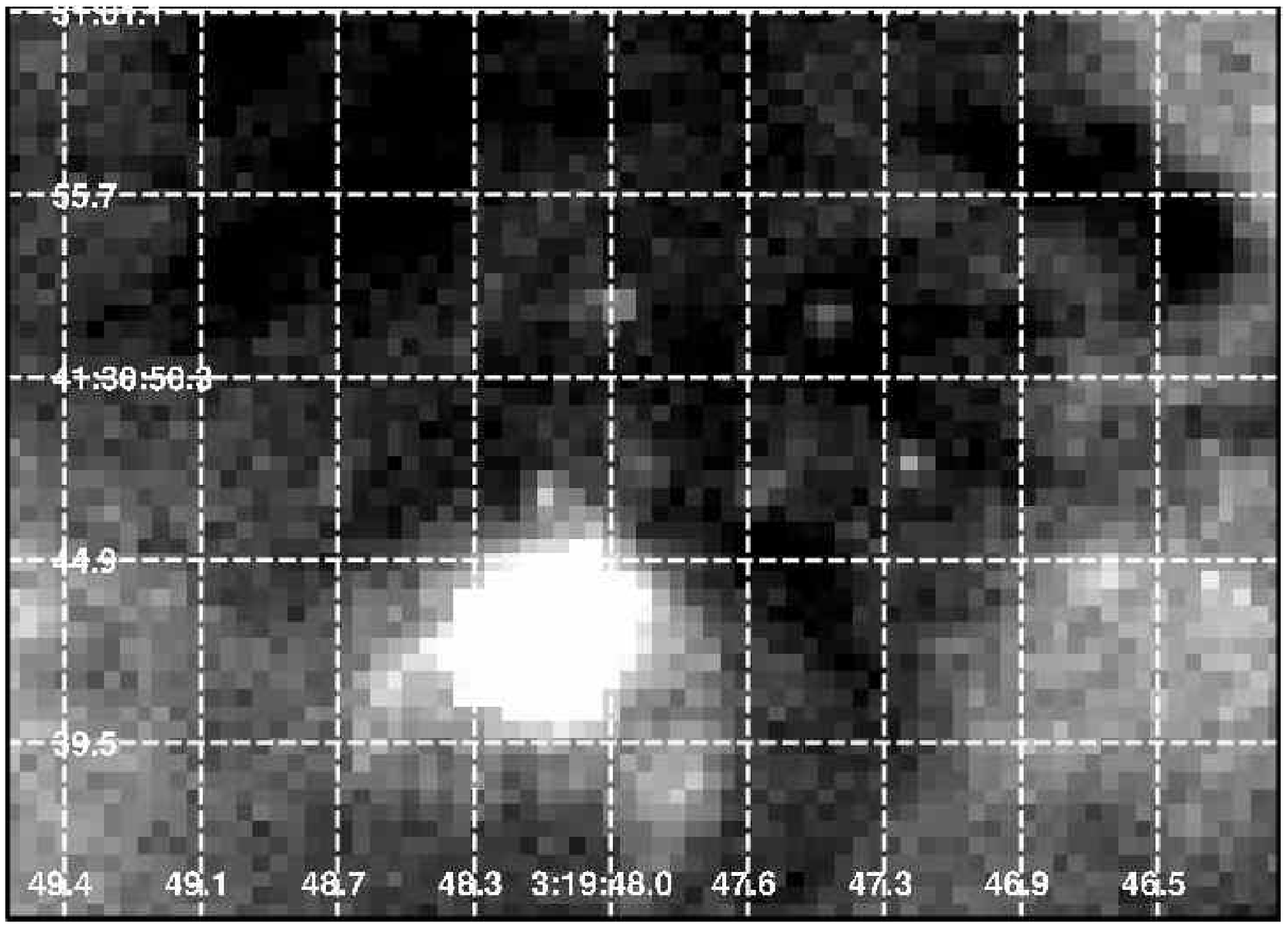}
 \includegraphics[width=\columnwidth]{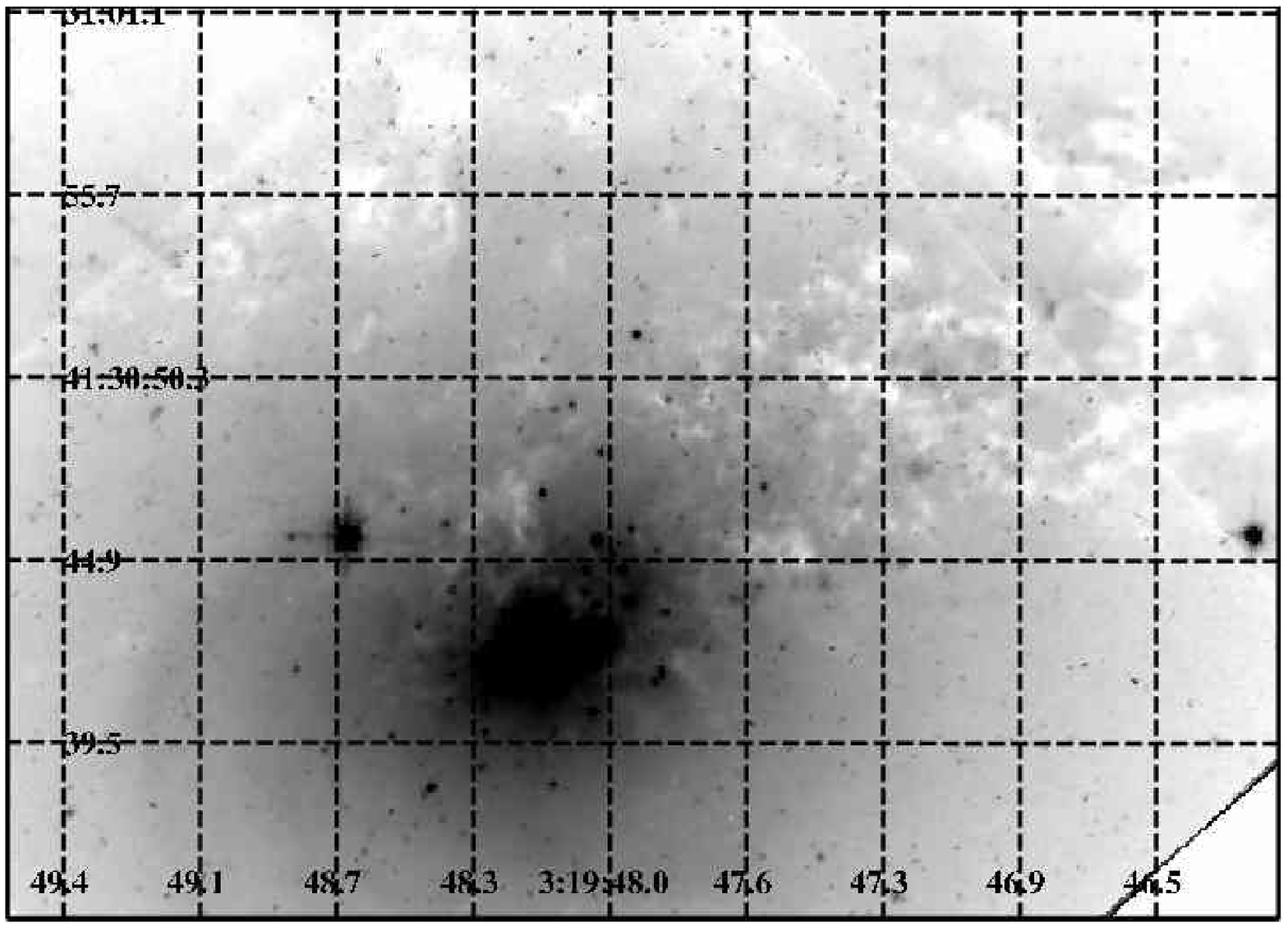}
  \caption{
    Top-left: Broadband (0.5--7.0 keV) X-ray smoothed image with sources labelled.
    Top-right: \emph{HST}/WFPC2 F814W broadband image.
    Centre-left: Broadband (0.5--7.0 keV) X-ray smoothed image centred in
    source 3.
    Centre-right: \emph{HST}/WFPC2 F814W broadband image centred in source 3.
    Bottom-left: Broadband (0.5--7.0 keV) X-ray smoothed image centred in
    source 5.
    Bottom-left: \emph{HST}/WFPC2 F814W broadband image centred in source 5.}
  \label{fig:figure}
\end{figure*}

As an example of our spectral fits, the brightest source, N3, has been
fitted with a power-law with spectral index of $2.08\pm0.09$ and absorption of
$\rm{N_H=2.5\pm0.4 \times 10^{21}~cm^{-2}}$ (see Fig.
\ref{fig:source1_spec}).

 \begin{table}
\begin{center}
  \begin{tabular}{llll} \hline \hline 
N &  F$\rm{_{obs}}$(0.5--7.0 keV) &  F$\rm{_{corr}}$(0.5--7.0 keV) & $\rm{log~L_{X}}$  \\ 
 &  erg $\rm{cm^{-2}~s^{-1}}$   &  erg $\rm{cm^{-2}~s^{-1}}$  & 0.5--7.0 keV  \\ \hline  
1.... &   2.09 $\times 10^{-15}$  &   4.34 $\times 10^{-15}$ &39.51   \\
2.... &   7.59 $\times 10^{-15}$  &   9.97 $\times 10^{-15}$ &39.86    \\  
3.... &   1.64 $\times 10^{-14}$  &   2.28 $\times 10^{-14}$ &40.22	 \\  
4.... &   7.76 $\times 10^{-15}$  &   1.10 $\times 10^{-14}$ &39.91    \\  
5.... &   5.36 $\times 10^{-15}$  &   1.07 $\times 10^{-14}$ &39.90    \\  
6.... &   3.02 $\times 10^{-15}$  &   9.23 $\times 10^{-15}$ &39.84	\\  
7.... &   4.28 $\times 10^{-15}$  &   1.18 $\times 10^{-14}$ &39.95    \\  
8.... &   7.67 $\times 10^{-15}$  &   1.16 $\times 10^{-14}$ &39.93    \\    \hline 
\end{tabular}
  \caption{Fluxes (observed and k-corrected) and luminosities assuming a cosmological
  model with $\rm{H_{0}=70~km~s^{-1} Mpc^{-1}}$ and z=0.018.} 
  \label{tab:luminosities}
\end{center}
\end{table}

In Table \ref{tab:luminosities} we list the 0.5-7~keV flux and
(absorption corrected) luminosities of the individual sources based on
the best-fit power law model.

The lower limit of the luminosity of point sources in the image, if at
the distance of NGC\,1275, is $\rm{L_X(0.5-7.0~keV)=3.2\times
  10^{39}erg~s^{-1}}$, which is already well above the Eddington limit for
a neutron star binary ($\rm{L_X\sim 3 \times 10^{38}erg~ s^{-1}}$) and
is also above the limit of canonical ULX, i.e. $\rm{\ge 10^{39}erg ~
  s^{-1}}$.

The brightest point source has a luminosity of
$\rm{L_X(0.5-7.0~keV)=1.67 \times 10^{40} erg ~ s^{-1}}$, and is one
of the brightest individual sources found in a galaxy. A ULX source
more luminous than the entire X-ray luminosity of a normal galaxy has
been found in the Cartwheel system with a luminosity of at least
$\rm{L_X \sim 2-4 \times 10^{40} erg ~ s^{-1}}$ (Gao et al. 2003;
Wolter \& Trinchieri 2004). They explain this luminosity with a
high-mass X-ray binary source (HMXB). The high X-ray luminosity
suggests either a single extremely bright source, or a very dense
collection of several high $\rm{L_X}$ sources, which would be even
more peculiar.  Evidence of time variability might suggest that is a
single high $\rm{L_X}$ source.

Time variability analysis has been performed. The observations span
about two years. Two data files were observed on 2002 August 8 and 10,
and the other eleven data files were observed from 2004 October 4 to
2004 October 20, giving an almost daily coverage. The exposure times
are between 22 and 143 ks. The data characteristics allows us
determine short variation in 16 days (second period) and long-term
variability of 2 years.  Because of the low count rates of the sources
in NGC\,1275 (see Table \ref{tab:positions}), it is very hard to
search for short-term variability. We extracted light-curves, using
{\sc dmextract} CIAO task for the two brightest sources (N3 and N4)
(net count rate greater than 0.98 count $\rm{s^{-1}}$) binned with bin
sizes of 500, 1000, 2500 and 5000 s. In both cases the points were
consistent with the respective mean values and variability has not
been found. Furthermore, the mean values between 2002 and 2004 are the
same, including errors bars. Therefore, evidence of time variability
has not been found during the whole set of observations.

\section{Discussion}
\emph{Chandra} has revealed significant populations of ULX in the
interacting systems of the Antennae (NGC 4038/9; Zezas, Fabbiano \&
Murray 2002) and the Cartwheel ring galaxy (Gao et al. 2003; Wolter \&
Trinchieri 2004), where dramatic events have stimulated massive star
formation. We have reported here on another example (Fig.
\ref{fig:figure} left) in the HVS of NGC\,1275 which is interacting
with the ICM of the Perseus cluster.

The sources are spatially associated with the distribution of
absorbing clouds seen in soft X-ray (Fig.~2) and optical (Fig.~3)
images. Two sources (N7 and N8) are directly linked with dust knots
and another two (N2 and N6) have an optical point-like counterpart
(Fig. \ref{fig:figure} \emph{bottom}). Similar correspondence have
been found in the Cartwheel galaxy with the outer ring (Wolter \&
Trinchieri 2004) and in the Antennae galaxies with 39 X-ray sources
within the WFPC2 field (Zezas et al. 2002). The optical brightness of
the counterparts in the HVC are too high to be individual stars and so
may be associated with young star clusters. Following the discussion
of young star clusters in NCG\,1275 given by Richer et al (1993), an
object of magnitude 22 corresponds to a cluster mass of about
$10^6\Msun$ if its age is about $10^7\yr$. The HVC system travels at
least 30~kpc in $10^7\yr$ so if a strong interaction with the core of
the Perseus galaxy cluster has triggered star cluster formation in the
HVC, then the clusters should have ages less than $\sim 10^8\yr$. 
 
Our interpretation of the spatial correspondence with star clusters is
that the regions are especially active, indicating a real link between
ULX and star-forming regions, and meaning they are young objects.
However the optical limits on sources N3 and 4 rule out any
association with massive clusters in those cases (the limit on the
absolute magnitude is about $-8$).

In M31 and the Milky Way (Grimm, Gilfanov \& Sunyaev 2003), XRB have
luminosities consistent with the Eddington limit of a $\rm{\sim 2
M_\odot}$ accreting object. They produce luminosities $\rm{\sim 3
\times 10^{38}~ erg~s^{-1}}$, about one order of magnitude below the
limiting luminosity in our sample ($\rm{3.2 \times 10^{39}~
erg~s^{-1}}$). It is possible that our ULX consist of at least 15 (or
130, in the case of the brightest source found) `normal' XRB clustered
together, perhaps in a young star cluster. However in other objects we
know that variability requires the presence of intrinsically luminous
X-ray sources (e.g. M82; Griffiths et al. 2000, Kaaret et al. 2001).
Alternative possibilities are that black hole sources, with masses in
the range of galactic black hole binaries, are mildly beamed (Reynolds
et al. 1999 and King et al. 2001). Spectral and timing features
however rule out this possibility in some ULX (e.g. Strohmayer \&
Mushotzky 2003). We note that compact supernova remnants sometimes
have ULX luminosities (e.g. Fabian \& Terlevich 1998), but no recent
supernovae have been reported for NGC\,1275 (SN1968A was to the S of
the HVS; Capetti 2002).

Finally, we recall the IMBH model which has spectral support from some
sources (Miller et al 2004; the level of absorption in NGC\,1275 is
too high for any soft excess to be observed). They may form in dense
star clusters.

Our optical studies have clearly shown that the ULX have very high
X-Ray to optical flux ratios. X-ray selected AGN from the \emph{Rosat
all sky survey} tend to have $\rm{log(F_X/F_{opt})\sim 1}$. Thus the
ULX do not have the optical properties expected if their were simple
extensions of AGN (IMBH, as low luminosity limit). However, low mass
X-ray binaries in the Milky Way have $\rm{F_X/F_{opt}\sim 100-10000}$
(Mushotzky 2004).

The results found in our system indicate that we have a mixed group of
objects (see Table \ref{tab:optical}). At least 4 out of 8 sources
(N3, N4, N5 and N8) have high X-ray to optical flux ratios. At
least 3 out of 8 (N1, N2 and N6) have lower X-ray to optical ratios,
possibly because they lie in star clusters.

Our data are consistent with no significant variability, similar to
the result obtained on NGC\,3256 by Lira et al. (2002). Time
variability is frequently observed in ULX (e.g. IC\,342, Sugiho et al.
2001 or M51 X-1, Liu et al. 2002), arguing that most of them are
single compact objects, rather than a sum of numerous lower luminosity
objects in the same object. While most ULX vary, many show low
amplitude variability on long time scales (e.g. the Antennae galaxies,
Zezas et al. 2002), which is very different to galactic black holes.
Portegies Zwart, Dewi \& Maccarone (2004) find that a persistent
bright ULX requires a doner star exceeding $15\Msun$.  The search for
characteristic frequencies is one of the most productive way of
determining the nature of the ULX.

\section{Conclusions}
We have described the detailed analysis of the spatial and spectral
properties of the discrete X-ray sources detected with a deep
\emph{Chandra} ACIS-S observation around NGC\,1275. Our results are
summarized below:

\begin{enumerate}
\item We have detected a total of 8 sources to the north of NGC\,1275
  nucleus.
\item The sources are spatially coincident with the High Velocity
  System and thus probably associated with it. They are therefore ULX.
\item Four of the sources have an optical counterpart in the I and R
  bands (from \emph{HST} images); two of which are point-like sources and the
  other two are associated with star-forming regions.
\item In all the cases a single component power law gives satisfactory
  fits, with spectral index of $\rm{\Gamma=}$[1.78-3.51] and an
  equivalent column density of $\rm{N_H=[2.05-4.03]\times
    10^{21}cm^{-2}}$.
\item The minimum luminosity is $\rm{L_X(0.5-7.0 keV)=3.2\times
    10^{39}erg~ s^{-1}}$ (source N1), which is already above the limit
  of canonical ULX.
\item No variability was detected in the two brightest sources found.
\end{enumerate}

Our results add to the growing evidence that some episodes of rapid
star formation lead to the production of ULX.  Young, massive, star
clusters may be involved in some, but not all of the sources.

\section*{Acknowledgements}
OGM acknowledges the financial support by the Ministerio de Educacion
y Ciencia through the program AYA2003-00128 and grant FPI
BES-2004-5044. ACF thanks the Royal Society for support.

\end{document}